# Temperature and magnetic field induced structural transformation in Si doped CeFe$_2$: in-field x-ray diffraction study

Arabinda Haldar,[1] Niraj K. Singh,[2] Ya. Mudryk,[2] K. G. Suresh,[1,*] A. K. Nigam,[3] and V. K. Pecharsky[2,4]

[1]Department of Physics, Indian Institute of Technology Bombay, Mumbai-400076, India
[2]The Ames Laboratory U. S. Department of Energy, Iowa State University, Ames, Iowa 50011-3020, USA
[3]Tata Institute of Fundamental Research, Homi Bhabha Road, Mumbai 400005, India
[4]Department of Materials Science and Engineering, Iowa State University, Ames, Iowa 50011-2300, USA

*Abstract*

Using x-ray powder diffraction technique at various temperatures and applied magnetic fields, we have studied the magnetostructural properties of Ce(Fe$_{0.95}$Si$_{0.05}$)$_2$. The x-ray diffraction data establish quantitative relationships between bulk magnetization and the evolution of structurally distinct phases with magnetic field and temperature, and confirm the distinct features of first order phase transition like supercooling and superheating, metastability, and phase co-existence of different structural polymorphs. We observe the lattice volume mismatch across the structural phase transition, which appears to be the cause for the step behavior of the magnetization isotherms at low temperatures. The present study shows that the lattice distortion has to be treated explicitly, like spin, along with the effects of lattice-spin coupling to account for the magnetization behavior of this system. This structure template can resolve the issue of kinetics in this material as observed in different time scale measurements and with different experimental protocols.

[*]Author to whom any correspondence should be addressed (E-mail: suresh@phy.iitb.ac.in).

## 1. Introduction

Occurrence of a structural transformation along with a magnetic phase transition observed in some materials leads to interesting physics and underlies many unusual properties [1-5]. Anomalous behaviors of magnetization and resistivity obtained in certain manganites and a few intermetallics have been attributed to the field-induced magnetostructural transitions [1-9]. Martensitic scenario which arises from the first order structural distortion has been recently established in different classes of compounds [6-9]. Distinct features of first order phase transition have been illustrated by probing the magnetism of such materials both using bulk techniques [10] as well as microscopic (local) probes [11].

Magnetic and structural phase coexistence and metastability of magnetic phases across the transition have been experimentally investigated in a few $CeFe_2$ based compounds [11]. Sharp metamagnetic steps in the magnetization isotherms associated with structural changes were observed in different classes of materials, and have attracted a lot of attention because of the universality of this phenomenon [12-14]. Based on the magnetostriction and neutron diffraction studies it was proposed that step behavior seen in the low temperature magnetization isotherms is linked to the catastrophic relief of strain build up during the first order magnetostructural phase transition in which the magnetic state changes from AFM to FM [15]. The martensitic strains are a result of the lattice volume mismatch between the two different crystallographic structures corresponding to the AFM and FM phases. However, no systematic studies of both magnetization and crystal structure as a function of magnetic field and temperature have been reported to date in any of the doped $CeFe_2$ compounds.

Anomalous behavior in doped $CeFe_2$ compounds has attracted attention of many researchers for a long time [16-18]. Certain substitutions such as Ru, Re, Ga and Si are known to stabilize the fluctuating low temperature antiferromagnetism (AFM) of $CeFe_2$ [19]. The high temperature ferromagnetic (FM) phase is known to change to a low temperature antiferromagnetic phase on cooling. This is accompanied by a crystallographic distortion from cubic to rhombohedral structure [16, 20]. Magnetic phase transition and co-existence of magnetic phases have been studied by different experimental means. It has been found that the features associated with the martensitic scenario in the manganites and the intermetallic compounds such as $Gd_5Ge_4$ and doped $CeFe_2$ have many similarities [8, 9]. Recently, Ahn *et al.* have shown from their theoretical model that the micrometer-scale multiphase co-existence is self-organized and originates from the intrinsic lattice degrees of freedom rather than the charge density in phase separated manganites [1]. Therefore, it is important to analyze the structural variation with field and temperature in order to establish the origin of the distinct features like phase co-existence, supercooling and superheating, jumps in magnetization and resistivity, etc. in intermetallic compounds. Doped $CeFe_2$ compounds offer an ideal avenue for such a study.

Recently, we have found that Ga substitution stabilizes the AFM state in $CeFe_2$ and gives rise to the above mentioned features associated with first order transition [9]. Similar martensitic behavior has been found in Si doped compounds as well [21]. In this work, we have investigated the magneto-structural evolution in $Ce(Fe_{0.95}Si_{0.05})_2$ system using the in situ x-ray powder diffraction study at various temperatures and magnetic fields across the FM-AFM phase transition.

## 2. Experimental details

Polycrystalline $Ce(Fe_{0.95}Si_{0.05})_2$ compound was prepared by methods reported elsewhere [21]. The temperature (10-295 K) and field (0-40 kOe) dependent x-ray powder diffraction (XRD) data were collected using a Rigaku TTRAX powder diffractometer with Mo $K_\alpha$ radiation [22]. In view of the strong oxidation tendency of $Ce(Fe_{0.95}Si_{0.05})_2$, the fine powder (<25 μm) of the sample was prepared in a glove box. The powder was then mixed with GE varnish and dried in air for 4 days in order to solidify the specimen and prevent the rotation of individual particles by the magnetic field. To reduce surface roughness and preferred orientation, a flat surface was created using a 400 grit sandpaper. Multiple sets of diffraction data were collected in step (0.5–2 s/step) scanning mode, with a 0.01° step of $2\theta$ over the range of 9° ≤ $2\theta$ ≤ 45°. Each data set was analyzed by Rietveld refinement to determine the unit cell dimensions and phase contents, when two different crystallographic phases coexisted in certain field and temperature regimes. Temperature and field dependencies of magnetization data have been collected in zero field cooling (ZFC), field cooled cooling (FCC) and field cooled warming (FCW) protocols using a vibrating sample magnetometer attached to a commercial PPMS (Quantum Design).

## 3. Results and discussions

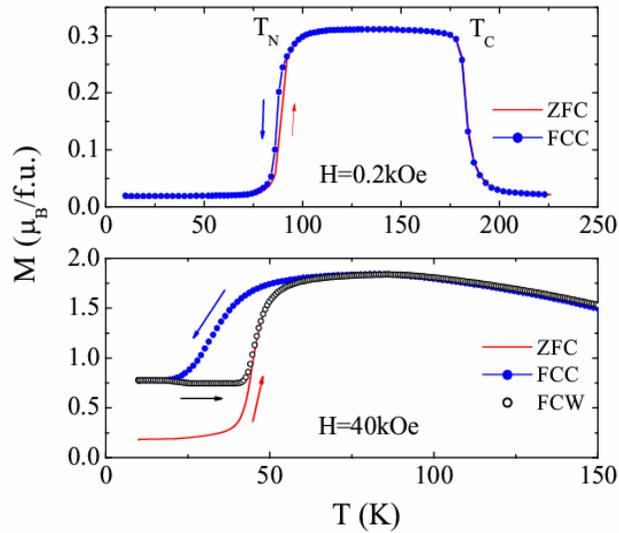

**Fig. 1.** Temperature variation of magnetization in ZFC, FCC and FCW modes in Ce(Fe$_{0.95}$Si$_{0.05}$)$_2$ at H=0.2 kOe and 40 kOe.

Fig. 1 shows the temperature variation of magnetization data in ZFC, FCC and FCW modes of Ce(Fe$_{0.95}$Si$_{0.05}$)$_2$ in 200 Oe and 40 kOe magnetic fields. As reported earlier, with Si substitution, the fluctuating low temperature antiferromagnetic ground state in undoped CeFe$_2$ gets stabilized [21]. The compound shows a paramagnetic (PM) – ferromagnetic transition during heating/cooling at the Curie temperature (T$_C$) of 184 K. During cooling it undergoes a FM-AFM transition at T$_N$=87 K. On warming, the AFM-FM transition occurs at 90 K. The thermal hysteresis observed between the cooling and warming data across the AFM-FM transition region is in agreement with earlier reports on Ru-doped CeFe$_2$ and indicates the first order nature of the transition[11]. As a consequence of the first order transition, it has been found that a fraction of the high temperature structural/magnetic phase gets supercooled to temperatures below $T_N$, thereby causing the coexistence of the high and low temperature phases[11]. It can be seen from Fig. 1 that the bifurcation between the FCC and ZFC magnetization data at the AFM-FM transition region remarkably increases when the field is increased to 40 kOe. This is due to the fact that the higher applied field enhances the supercooling of the FM component below $T_N$. The magnetization data strongly indicate the possibility of magnetostructural coupling at AFM-FM transition. Therefore, it is important to find out the evolution of structural phases with temperature in various applied fields, with different measurement protocols.

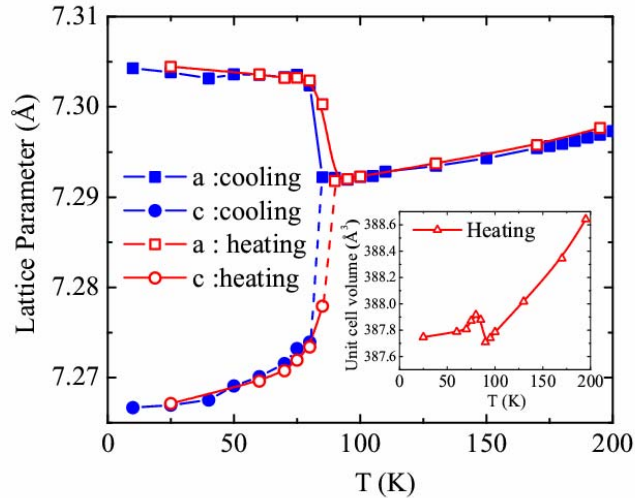

**Fig. 2.** Temperature variation of the normalized lattice parameters in $Ce(Fe_{0.95}Si_{0.05})_2$ during cooling and heating in zero magnetic field. Inset shows the variation of the unit cell volume with temperature.

The temperature-dependent x-ray powder diffraction data reveal that the $Ce(Fe_{0.95}Si_{0.05})_2$ possesses cubic Laves phase structure at room temperature and the cubic structure is preserved down to 90 K during cooling. On further cooling, the compound exhibits a structural transformation from the cubic to rhombohedral structure at 85 K and the rhombohedral structure is retained down to 10 K. Fig. 2 shows the temperature variation of normalized lattice parameters obtained from the Rietveld refinement of x-ray diffraction data collected at various temperatures during cooling and heating modes in zero field. The lattice parameters of the rhombohedral phase were calculated by Rietveld refinement using hexagonal setting of the R3m space group symmetry. The calculated $a_{rh.}$ and $c_{rh.}$ parameters were modified ($a_{rh.}$ was multiplied by $\sqrt{2}$, and $c_{rh.}$ was divided by $\sqrt{3}$) in order to be directly compared with the high-temperature cubic lattice parameter. The unit cell volume was also normalized: $V = V_{rh.} \times 4/3$. During warming, the reverse transformation from the rhombohedral to cubic structure occurs at 90 K.

The structural transition temperature during cooling (heating) coincides with the magnetic transition temperature from FM (AFM) phase to AFM (FM) phase. Therefore, the temperature dependent x-ray data reveal that in the AFM phase the $Ce(Fe_{0.95}Si_{0.05})_2$ possesses rhombohedral structure, whereas in the FM and the PM phases it possesses cubic structure. The thermal irreversibility associated with the structural phase transformation may arise due to limitations to the nucleation and growth of one phase at the expense of the other during the structural transformation because of the lattice strain built-up. From Fig. 1 and 2, it is clear that the cubic (FM) - rhombohedral (AFM) transition during cooling occurs at a lower temperature compared to the rhombohedral (AFM) - cubic (FM) transition during warming as a result of superheating and supercooling effect [23]. It may be noted from the inset of Fig. 2 that the temperature variation of unit cell volume associated with the magnetostructural transition shows a discontinuity and the unit cell volume of the rhombohedral phase is higher compared to that of the cubic phase.

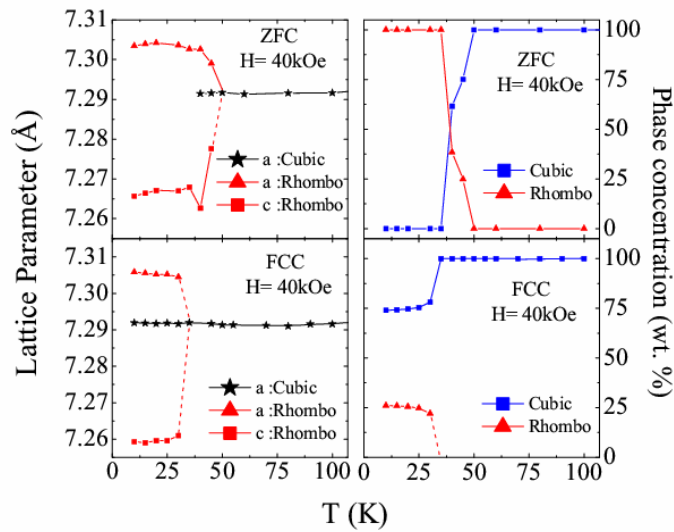

**Fig. 3.** (Left panel) Temperature variation of the normalized lattice parameter in ZFC and FCC modes at H=40 kOe. Right panel shows the concentrations of phases with different crystal structures as a function of temperature.

In order to further understand the magnetostructural properties of Ce(Fe$_{0.95}$Si$_{0.05}$)$_2$, we have also studied the temperature dependence of the crystal structure using x-ray powder diffraction in an applied magnetic field (H) of 40 kOe under ZFC and FCC protocols and the results are shown in Fig. 3. We note that in the ZFC mode, in 10-35 K temperature range, the compound is fully in the rhombohedral structure. However, phase co-existence is observed across the rhombohedral to cubic phase transition region. Furthermore, it may be noted that while 40 kOe field is insufficient to cause the rhombohedral to cubic transformation at low temperatures in ZFC protocol, in the FCC protocol the cubic phase coexists with the rhombohedral phase over a large temperature span. The phase coexistence is attributed to the supercooling effect, which facilitates the formation of FM phase at the expense of the AFM phase with increase in the applied field. We recall here that the M-T data (see Fig. 1) show that under the ZFC protocol, at low temperatures, the magnetization of Ce(Fe$_{0.95}$Si$_{0.05}$)$_2$ is substantially lower compared to the magnetization under FCC protocol, which indicates the presence of the high temperature ferromagnetic phase at low temperatures under FCC protocol. This observation corroborates the x-ray results and indicates that in Ce(Fe$_{0.95}$Si$_{0.05}$)$_2$ the magnetism and structure are intimately coupled.

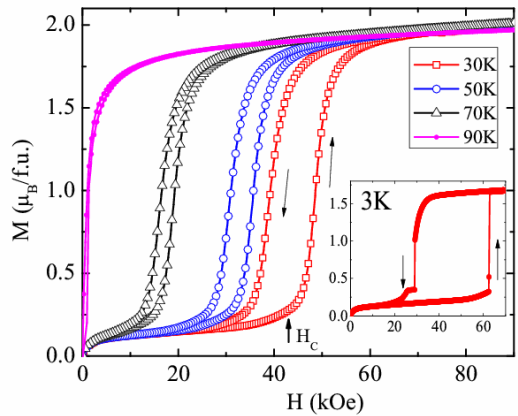

**Fig. 4.** Magnetization isotherms of Ce(Fe$_{0.95}$Si$_{0.05}$)$_2$ at 30, 50, 70 and 90 K. Sample was zero field cooled from above its T$_C$ between successive measurements. Inset shows the magnetization isotherm taken at 3K for increasing and decreasing fields.

Fig. 4 shows the field dependence of magnetization of Ce(Fe$_{0.95}$Si$_{0.05}$)$_2$. Similar observations have been reported for other doped CeFe$_2$ compounds [9,15]. We note that below T$_N$ the compound shows metamagnetic transition from AFM to FM phase. Here we define critical field H$_C$ as the field at which the slope change occurs drastically in the magnetization isotherms, across metamagnetic transition (from AFM to FM). H$_C$ is different for the increasing and decreasing field cycles. Even more interestingly, field hysteresis increases when the temperature decreases, meaning that the "freezing" boundary widens. The energy barrier for the FM-AFM return magnetic transition is, probably, higher at low temperatures.

On further reducing the temperature, the M(H) isotherm changes dramatically (see inset of Fig. 4). At 3 K, magnetization shows a huge jump while transforming from AFM to FM phase. Therefore, it is clear that large thermal fluctuations cause a comparatively smooth transition across the metamagnetic region at high temperatures. The irreversibility between the field up and field down data is attributed to the supercooling and superheating of the ferromagnetic phase during field cycling. In CeFe$_2$-based materials similar effects of supercooling and superheating have been observed with Hall probe measurement [11]. In the light of these observations, it is of importance to examine how the structure transforms with the application of the magnetic field.

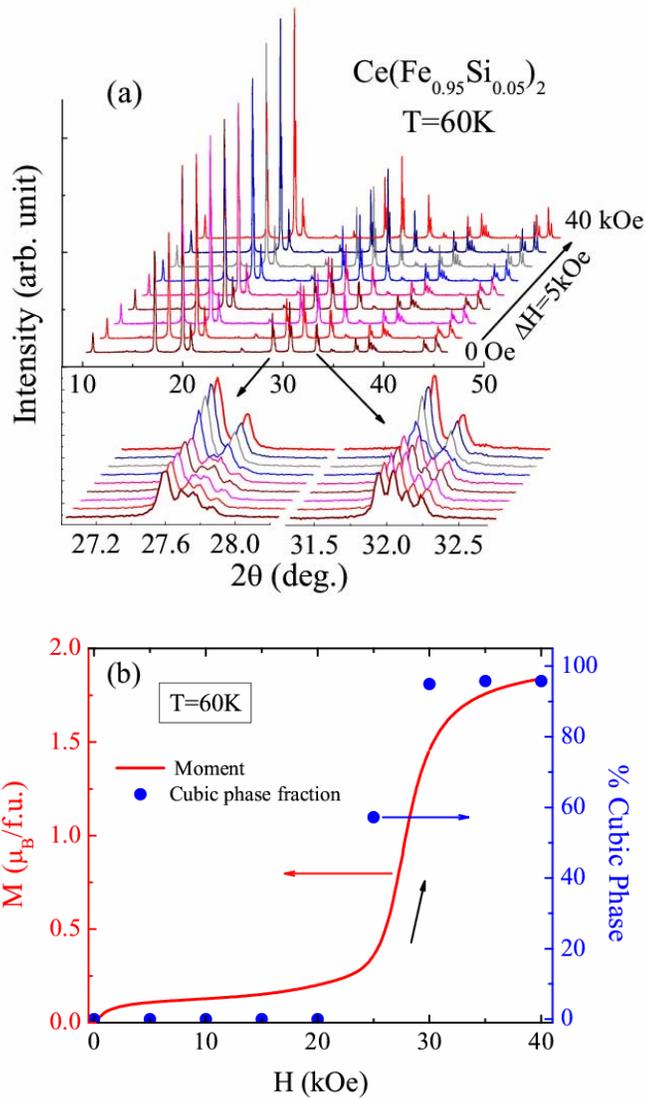

**Fig. 5.** (a) Field variation of x-ray powder diffraction patterns of Ce(Fe$_{0.95}$Si$_{0.05}$)$_2$ at 60 K. Lower panel shows the zoomed portion of two characteristic regions of the pattern, (b) Variation of isothermal magnetization and the percentage of the cubic (FM) phase as a function of applied field at 60K.

Fig. 5(a) shows the field dependence of the XRD data of Ce(Fe$_{0.95}$Si$_{0.05}$)$_2$ collected at 60 K. The measurement has been done with a field step of 5 kOe. The compound was cooled from 305 K to 60 K in zero field and the x-ray diffraction patterns were recorded while increasing the field. In order to highlight the effect of the magnetic field on the crystal structure, in the lower panel of Fig. 5(a), two selected regions (27º-28.3º and 31.3º-32.8º) are shown in the expanded view. We note that around 27.7º and 32º, four Bragg reflections of the rhombohedral phase ( [214] and [018] for 27.7°, and [220] and [208] for 32°) merge into two Bragg peaks of the cubic phase ([224] and [044], respectively) with increasing field, which indicates the rhombohedral to cubic crystallographic phase change.

The M-H isotherm and the calculated phase fraction of the cubic (FM) phase with varying field at 60 K are shown in Fig. 5(b). It is found that at 60 K the rhombohedral structure is retained up to the field of 20 kOe. However, at 25 kOe, more than 50% of the rhombohedral phase is converted to the cubic phase. It is interesting to note that the ZFC M(H) of Ce(Fe$_{0.95}$Si$_{0.05}$)$_2$ also shows the metamagnetic phase transition at the same field. With further increase in field ($\geq 30 kOe$) almost all (~95%) rhombohedral phase is converted into the cubic phase, thus our field dependent x-ray diffraction data show that in this metamagnetic process the concentration of the cubic polymorph of the compound is in a quantitative relationship with the evolution of bulk magnetization.

The agreement between the growth of the net magnetization and the percentage of the cubic (FM) phase as a function of field is obvious. This observation serves as a strong experimental evidence for the expected coupling of lattice and spin phenomena in this compound. Moreover it is justified to conclude that the properties like supercooling/superheating, phase co-existence, etc. are associated with the structural distortion. So, we believe that the structural distortion has to be treated explicitly as a main contribution to the anomalous behavior of this, and probably other doped CeFe$_2$ compounds, along with its coupling with spins. Inclusion of lattice degrees of freedom, together with the lattice–spin coupling can explain the universality of the features observed in CeFe$_2$-based series of materials. Such a scenario is true in the case of phase separated perovskite manganites as was suggested earlier by Ahn *et al* [1]. Origin of steps in the magnetization isotherms at low temperatures in Ce(Fe$_{0.95}$Si$_{0.05}$)$_2$ can be explained with this clear picture of structural transformation. Volume mismatch (see inset of Fig. 2) between the two structures, rhombohedral and cubic, produces strains in the system. Relief of strains along with the moment reorientation from AFM to FM configuration makes the system convert to its high field stable phase in a burst-like fashion, resulting in the steps.

The above mentioned results help to explain another important feature generally seen in doped $CeFe_2$, which is the magnetic field sweep rate dependence of magnetization [9]. Due to the presence of relaxation in these materials, the magnetization behavior is found to depend on the experimental time scale of the measurement. It was found that the critical fields, at which the steps occur, decrease with increase in the field sweep rate. Furthermore, the number of steps is also dependent on the sweep rate. So it is important to investigate the kinetics of the magnetization process in this system. The issue of kinetics can be addressed under the template of structural transformation. It is to be noted here that each structural transformation possesses a characteristic relaxation time, which becomes sluggish with the decrease in temperature as a result of the reduction in the displacive motion of the atoms [24]. If the experimental time scale is not sufficient to convert a structure to its equilibrium phase at a certain temperature and magnetic field, the structural evolution will be slow. Due to the magnetostructural coupling, the moment relaxation will also be slow in such a case. It has been reported that such relaxation behavior is shown by some phase separated manganites [25, 26] as well as doped $CeFe_2$ [9].

## 4. Conclusions

To summarize, we have systematically studied the variation of crystal structure under ZFC, FCC and FCW modes in zero and high (40 kOe) magnetic fields in $Ce(Fe_{0.95}Si_{0.05})_2$. Temperature and field induced structural transformation has been observed from the XRD study. This, in conjunction with the magnetization results, confirms the presence of distinct features associated with the first order phase transition such as supercooling and superheating, metastability, and phase co-existence of different structural polymorphs. We have also shown the lattice volume mismatch across the structural phase transition, which is responsible for creating the martensitic strains in the AFM-FM transition region. Based on the present study, we conclude that lattice distortion, like spin, has to be taken into account along with the effects of lattice-spin coupling in order to understand the magnetization dynamics of doped CeFe$_2$. This structure template can resolve the issue of kinetics in this system as observed in different time scale measurements and with different experimental protocols.

*Acknowledgements*

KGS and AKN thank BRNS (DAE) for the financial assistance for carrying out this work. The Ames Laboratory is supported by the Office of Basic Energy Sciences, Materials Sciences Division of the U.S. Department of Energy under contract No. DE-AC02-07CH11358 with Iowa State University. The authors thank Prof. K.A. Gschneidner, Jr. for fruitful discussions.